\documentclass{PoS}

\usepackage{amsmath}
\usepackage{graphicx}
\usepackage{subfigure}

\newcommand{\diag}{{\rm diag\,}}
\newcommand{\Tr}{{\rm Tr\,}}
\newcommand{\rmR}{{\rm R}}
\newcommand{\rmI}{{\rm I}}

\title{Complex Langevin dynamics: criteria for correctness}

\ShortTitle{Complex Langevin dynamics: criteria for correctness}

\author{Gert Aarts\\
  Department of Physics, Swansea University, Swansea, UK\\
  E-mail: \email{g.aarts@swan.ac.uk}}

\author{
  \speaker{Frank A.\ James}\\
  Department of Physics, Swansea University, Swansea, UK\\
  E-mail: \email{pyfj@swan.ac.uk}}

\author{
  Erhard Seiler\\
  Max-Planck-Institut f\"ur Physik (Werner-Heisenberg-Institut), M\"unchen, Germany\\
  E-mail: \email{ehs@mppmu.mpg.de}}

\author{
  Ion-Olimpiu Stamatescu\\
  Institut f\"ur Theoretische Physik, Universit\"at Heidelberg and FEST, Heidelberg, Germany\\
  E-mail: \email{I.O.Stamatescu@thphys.uni-heidelberg.de}}

\abstract{The complex Langevin method is a leading candidate for solving the sign problem occurring in various
physical situations, notably QCD at finite chemical potential. Its most vexing problem is `convergence to the wrong
limit', where the simulation gives a well defined, but incorrect, result. Here, we first outline a formal justification
of the method and identify points at which it might fail. From these we derive a condition that must be satisfied in order
for correct results to be obtained.
We then apply these ideas to the three-dimensional SU(3) spin model at finite chemical potential and show strong indications that complex Langevin dynamics
yields correct results in this theory.}

\FullConference{XXIX International Symposium on Lattice Field Theory \\
  July 10 - 16 2011\\
  Squaw Valley, Lake Tahoe, California}

\begin{document}

\section{Introduction}
Many systems of physical relevance have a complex action, QCD with a finite chemical potential is a particularly interesting example. Such theories
are difficult to study numerically because the weight is complex and therefore standard Markov chain Monte Carlo techniques
based on a probability interpretation cannot be applied; this is often called the sign problem~\cite{deForcrand:2010ys}.
The complex Langevin equation (CLE) enjoyed a certain limited popularity after being proposed in the early 1980s~\cite{Parisi:1984cs, Klauder:1983zm} because it can, in principle, avoid the sign problem since it does not rely on importance sampling.
In some cases, complex Langevin simulations have been found to be numerically unstable due to runaway solutions but this
can be cured by use of an adaptive stepsize~\cite{Aarts:2009dg}.
A more serious and vexing issue is convergence to the wrong limit, where the simulation
gives well defined results, but when compared with known values are found to be incorrect. It is this problem that
we shall focus on here.

In Section~\ref{sec:cc} we briefly cover a formal argument for the correctness of the CLE and identify points at which
it might fail. By studying the long time evolution of observables with respect to real and complex measures,
we derive a criterion which must be satisfied in order for correct results to be obtained~\cite{Aarts:2009uq, Aarts:2011ax}. 
In Section~\ref{sec:su3} these ideas are applied to the SU(3) spin model at finite chemical potential~\cite{Karsch:1985cb, Bilic:1987fn, Gattringer:2011gq}.
A comparison with imaginary chemical potential
shows that the CLE works correctly in the region $\mu^2 \sim 0$. An analysis of the criterion with a larger chemical potential
supports the claim that in this case complex Langevin dynamics can be trusted~\cite{aarts-james}.

\section{Criteria for correctness\label{sec:cc}}
The central object of interest is the expectation value of a particular observable, given by
\begin{equation}
  \langle O\rangle = \frac{\int O(x) e^{-S(x)} dx}{\int e^{-S(x)}dx} ,
  \end{equation}
where for notational simplicity we use a single real degree of freedom, $x$. The action $S(x)$ is complex, preventing a probability interpretation
of the measure and ruling out methods based on importance sampling. 
The complex Langevin equation is
\begin{equation}
  \frac{dx}{dt} = K_x + \sqrt{N_\rmR}\eta_\rmR, \quad\quad\quad\quad \frac{dy}{dt} = K_y + \sqrt{N_\rmI}\eta_\rmI, 
\end{equation}
where the real variable is analytically continued as $x\to z = x+iy$. The drift terms are given by
\begin{equation}
  K_x = {\rm Re\,} \left. \frac{dS(x)}{dx}\right|_{x\to x+iy},\quad\quad\quad\quad K_y = {\rm Im\,} \left.\frac{dS(x)}{dx}\right|_{x\to x+iy},
\end{equation}
and the two noise terms $\eta_\rmR$, $\eta_\rmI$ are independent Gaussian random numbers with variance $2$ and normalisation $N_\rmI \ge 0$ and $N_\rmR - N_\rmI = 1$.
A numerical simulation can then be implemented by integrating these equations to large times $t\to\infty$. 

The resulting dynamics is described by a dual Fokker-Planck equation for the evolution of the probability density $P(x,y;t)$,
\begin{equation}
  \frac{\partial}{\partial t} P(x,y;t) = L^T P(x,y;t),
\end{equation}
with the operator 
\begin{equation}
  L^T = \nabla_x [N_\rmR \nabla_x - K_x] + \nabla_y [N_\rmI \nabla_y - K_y] .
\end{equation}

To understand the time evolution of the real density $P(x,y;t)$ one must also examine the evolution of the complex density $\rho(x;t)$, determined by
\begin{equation}
  \frac{\partial}{\partial t}\rho(x;t) = L^T_{0} \rho(x;t) .
  \label{eq:comp-density}
\end{equation}
Here, the complex Fokker-Planck operator $L^T_{0}$ is
\begin{equation}
  L^T_{0} = \nabla_x [\nabla_x + \nabla_x S(x)] .
  \label{eq:cfpe}
\end{equation}
This equation has $\rho(x;\infty) \propto \exp[-S(x)]$ as a stationary solution, which is expected to be unique.
Numerical studies (where feasible) of Eq.\ (\ref{eq:comp-density}) confirm this
to be true; in fact, convergence to this distribution seems exponentially fast.

Expectation values with respect to the two densities can now be defined as 
\begin{equation}
  \langle O\rangle_{P(t)} = \frac{\int O(x+iy)P(x,y;t) dxdy}{\int P(x,y;t) dxdy}, \quad\quad\quad\quad \langle O\rangle_{\rho(t)} = \frac{\int O(x)\rho(x;t) dx}{\int \rho(x;t)dx}.
\end{equation}
The result that one would like to show is
\begin{equation}
  \langle O\rangle_{P(t)} = \langle O\rangle_{\rho(t)},
  \label{eq:eql-exp-vals}
\end{equation}
if the initial conditions $\langle O\rangle_{P(0)} = \langle O\rangle_{\rho(0)}$ match,
which is assured provided
\begin{equation}
  P(x,y;0) = \rho(x;0)\delta(y).
\end{equation}
One expects the dependence on the initial conditions to vanish in the limit $t \to \infty$ by ergodicity. 
  
To establish a connection between the expectation values with respect to $P$ and $\rho$, one moves the time evolution
from the densities to the observables. Since we are only interested in functions of $z = x+iy$ (holomorphic functions), 
we may act with the Langevin operator
\[ \tilde{L} = [\nabla_z - (\nabla_z S(z))]\nabla_z ,\]
whose action on holomorphic functions agrees with that of $L$.

We now use $\tilde{L}$ to evolve observables according to the equation
\begin{equation}
  \frac{\partial}{\partial t}O(z;t) = \tilde{L}O(z;t) ,
\end{equation}
which is formally solved by
\[O(z;t) = \exp[t \tilde{L}] O(z).\]
Due to the fact that $L$ and $\tilde{L}$ agree on holomorphic functions, the tilde may be dropped.

To examine the evolution we define the function
\begin{equation}
  F(t,\tau) = \int P(x,y;t - \tau) O(x+iy;\tau) dxdy,
\end{equation}
and observe that $F(t,\tau)$ interpolates between the two expectation values:
\begin{equation}
  F(t,0) = \langle O\rangle_{P(t)}, \quad\quad\quad\quad F(t,t) = \langle O\rangle_{\rho(t)}.
\end{equation}
The first can be seen easily, while the second makes use of the initial conditions and 
\begin{align}
  F(t,t) &= \int P(x,y;0)\left(e^{t L}O\right)(x + iy;0) dxdy\nonumber\\
         &= \int \rho(x;0)\left(e^{tL_0}O\right)(x;0)dx = \int O(x;0)\left(e^{t L_0^T}\rho\right)(x;0)dx = \langle O\rangle_{\rho(t)},
\end{align}
where it is only necessary to assume that integration by parts in $x$ does not produce any boundary terms.

The desired result (\ref{eq:eql-exp-vals}) follows if $F(t,\tau)$ is independent of $\tau$. To check this, we need the
$\tau$ derivative to vanish,
\begin{equation}
  \frac{\partial}{\partial \tau} F(t,\tau) =- \int \left(L^T P(x,y;t-\tau)\right)O(x+iy;\tau) dxdy + \int P(x,y;t-\tau) LO(x+iy;\tau) dxdy.
  \label{eq:f-deriv}
\end{equation}
Integration by parts, \emph{if applicable without boundary terms at infinity},
then shows that the two terms cancel and therefore $F(t,\tau)$ is independent of $\tau$, irrespective of $N_\rmI$.

This is therefore a point at which the formal argument might fail: if the decay of the product
\[ P(x,y;t-\tau)O(x+iy;\tau) \]
and its derivatives is insufficient for integration by parts without boundary terms.

In Ref.~\cite{Aarts:2011ax} a study of the U(1) one-link model found that the $\tau$-derivative is largest at
$\tau=0$. This motivates the superficially weaker condition
\begin{equation}
  \left.\lim_{t\to \infty}\frac{\partial}{\partial t}F(t,\tau)\right|_{\tau=0} = 0.
\end{equation}
This modification to the condition is still sufficient for correctness if it holds for a sufficiently large
set of suitably chosen observables. Taking the limit $t\to\infty$ in Eq.\ (\ref{eq:f-deriv}) causes the first contribution
to vanish because of the equilibrium condition, $L^T P(x,y;\infty) = 0$. Therefore, the criterion for correctness reduces to
\begin{equation}
  E_O \equiv \int P(x,y;\infty) LO(x+iy,0) dxdy = \langle LO\rangle = 0.
  \label{eq:con-con}
\end{equation}
This is fairly simple to check for a given observable, but it is in fact a strong statement
since it must hold \emph{for all observables}. Therefore, Eq.\ (\ref{eq:con-con}) really represents an infinite tower
of identities which must all be satisfied. In practice it can be checked for a small number of observables, still yielding a necessary criterion~\cite{Aarts:2011ax}.
  
\section{SU(3) spin model \label{sec:su3}}
Motivated by recent work~\cite{Gattringer:2011gq} (for related models see e.g.\ Ref.~\cite{Lottini:2011bj}), we re-examine the three-dimensional SU(3) spin model at
finite chemical potential, for which promising results with complex Langevin dynamics have been obtained in earlier studies~\cite{Karsch:1985cb, Bilic:1987fn}.
The action is formed from three contributions
\begin{equation}
  S = S_B + S_F + S_H,
\end{equation}
which read 
\begin{align}
  S_B &= -\beta\sum_{x,\nu}\left( \Tr U_x \Tr U_{x+\hat\nu}^{-1} + \Tr U_x^{-1} \Tr U_{x+\hat\nu}\right),\\
  S_F &= -h\sum_x \left(e^\mu\Tr U_x + e^{-\mu}\Tr U_x^{-1}\right),\\
  S_H &= -\sum_x \log\left[\sin^2\left(\frac{\phi_{1x} - \phi_{2x}}{2}\right)
                           \sin^2\left(\frac{2\phi_{1x} + \phi_{2x}}{2}\right)
                           \sin^2\left(\frac{\phi_{1x} + 2\phi_{2x}}{2}\right)\right].
\end{align}
The final component, $S_H$, originates from the Haar measure introduced by diagonalising $U$ in terms of angles, 
\[ U = \diag (e^{i\phi_1}, e^{i\phi_2}, e^{-i(\phi_1 + \phi_2)}). \]

\begin{figure}[htb]
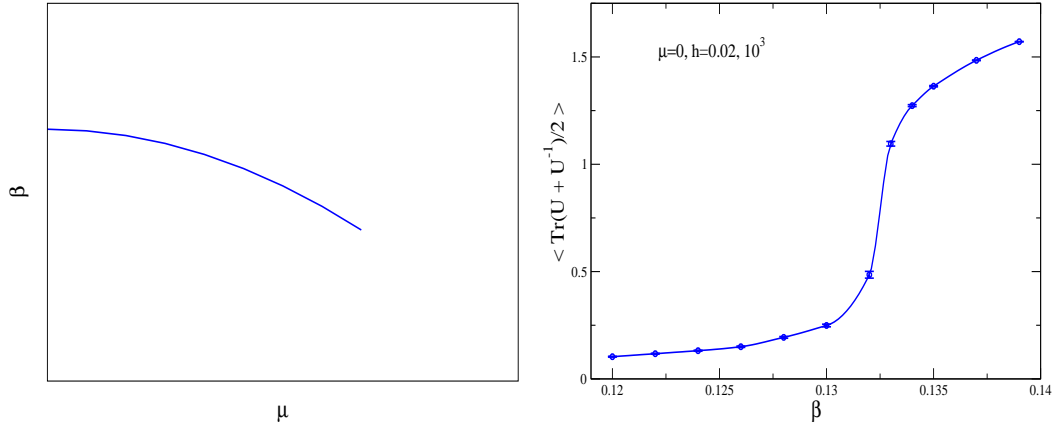

  \centering
  \subfigure{
    \includegraphics[width=0.45\textwidth, height=0.25\textheight]{plot_phasediagramKW.eps}
  }
  \subfigure{
    \includegraphics[width=0.45\textwidth, height=0.25\textheight]{plot_tru-invu_mu0_h0.02_10x10x10.eps}
  }
  \caption{Left: sketch of the phase diagram at small $h$. Right: $\langle \Tr (U + U^{-1})/2\rangle$ as a
  function of $\beta$ at $\mu=0$, $h=0.02$ on a $10^3$ lattice.}
  \label{fig:su3-pd}
\end{figure}

The action is complex due to the ``heavy fermion'' contribution, $S_{\rm F}^{*} (\mu) = S_{\rm F}(-\mu^{*})$.
For small $h$ the theory has a confined and deconfined phase separated by a first order transition line at small $\mu$,
turning into a crossover at larger $\mu$ (see Figure \ref{fig:su3-pd}, left)~\cite{Karsch:1985cb}.

\begin{figure}[!t]
  \centering
  \includegraphics[width=0.7\textwidth, height=0.4\textheight]{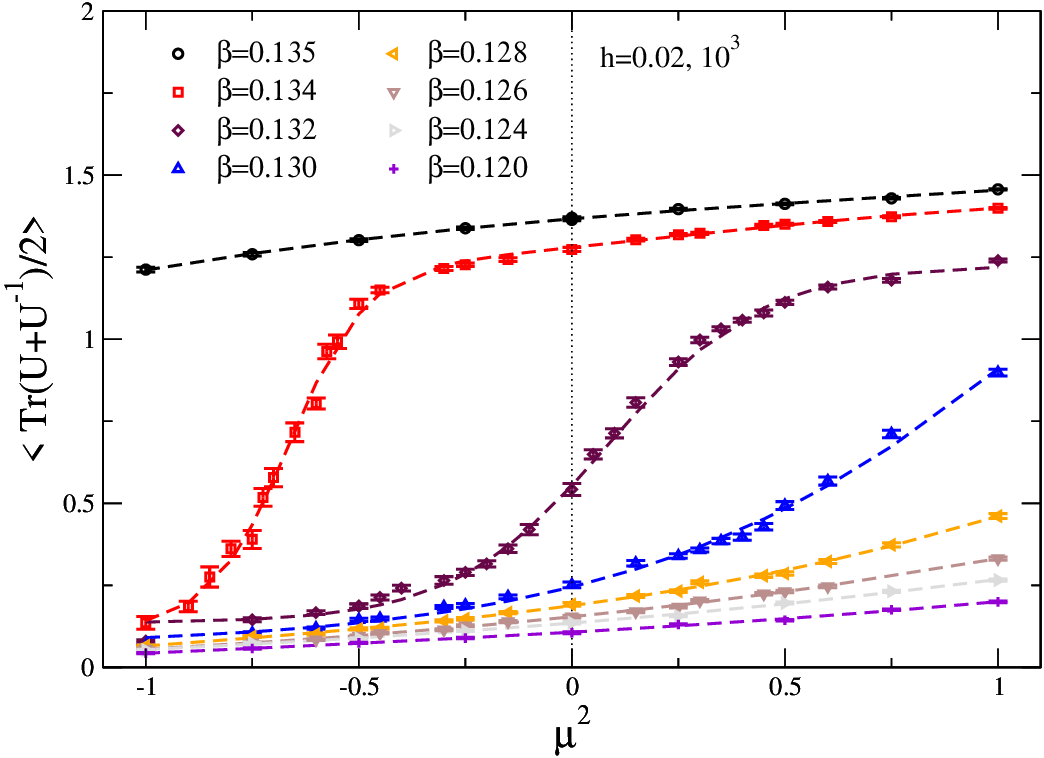}
  \caption{$\langle \Tr(U + U^{-1})/2\rangle$ as a function of $\mu^2$ for various $\beta$ values. For $\mu^2<0$ (imaginary $\mu$) the action
    is real and real Langevin dynamics is used; when $\mu^2 > 0$ complex Langevin is necessary.}
  \label{fig:su3-main-figure}
  \vspace{0.1in}
  \centering
  \subfigure[Complex Langevin dynamics]{
    \includegraphics[width=0.45\textwidth, height=0.25\textheight]{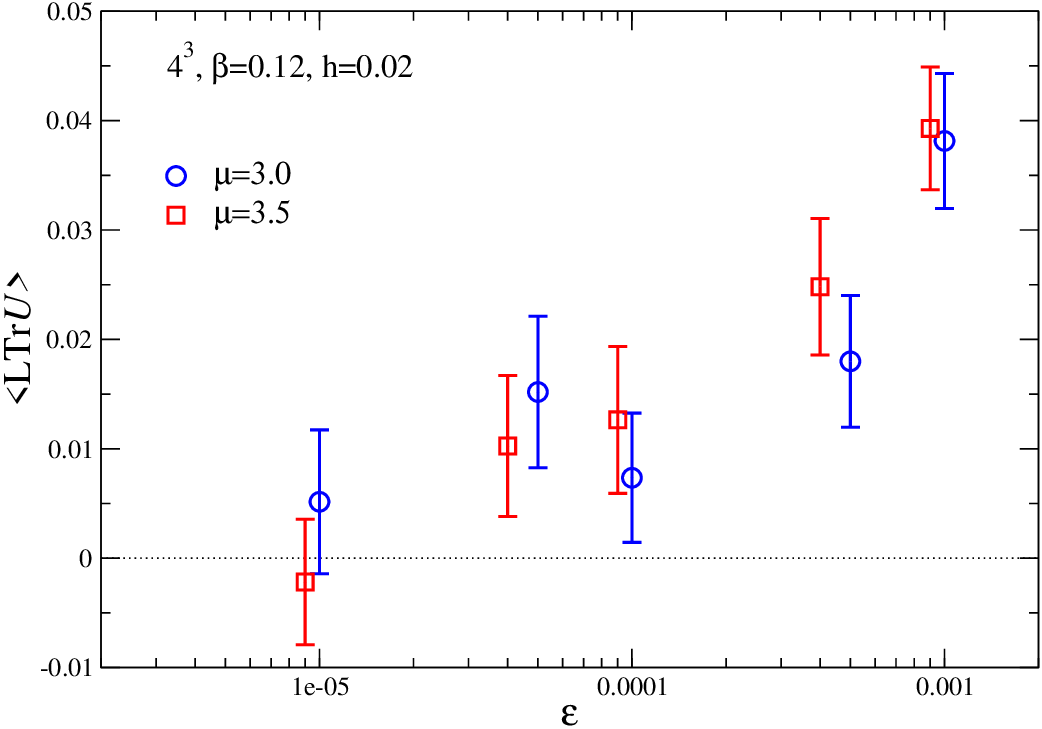}
  }
  \subfigure[Phase-quenched using real Langevin dynamics]{
    \includegraphics[width=0.45\textwidth, height=0.25\textheight]{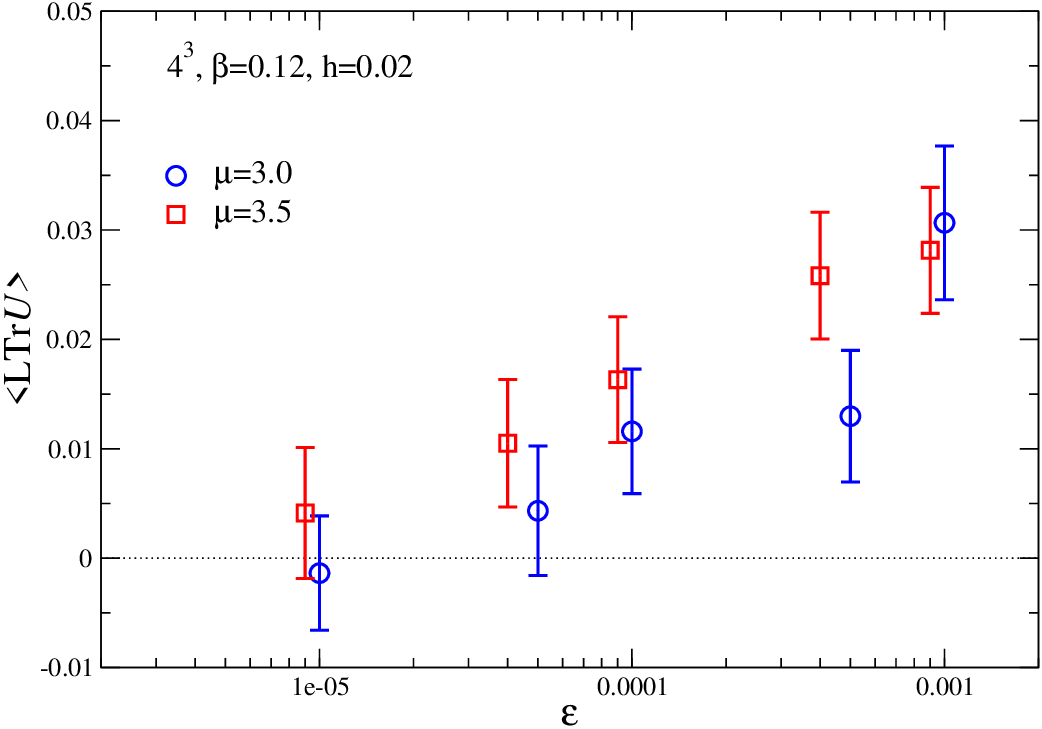}
  }
  \caption{Langevin stepsize dependence of the correctness criterion for $\Tr U$ with complex Langevin dynamics (left) and in the phase-quenched approximation (right)
    using real Langevin dynamics. The points for $\mu=3.5$ are shifted horizontally for clarity ($\beta=0.12$, $h=0.02$ and lattice volume $4^3$).}
  \label{fig:cc}
\end{figure}

Figure \ref{fig:su3-pd} (right) shows
the transition as a function of $\beta$ at $\mu=0$, using the observable $\langle \Tr (U+U^{-1})/2 \rangle$. Note that at
$\mu=0$, $\langle \Tr U\rangle = \langle \Tr U^{-1}\rangle$, but at $\mu^2 > 0$ they differ. 
Since observables that are even in $\mu$ should be continuous across $\mu^2=0$, we also perform simulations with
an imaginary chemical potential, for which the action is real and simulations using standard techniques can be used
(we use real Langevin dynamics). A comparison with complex Langevin data should therefore show even observables
to be continuous across $\mu^2=0$. In Figure~\ref{fig:su3-main-figure} the observable $\langle \Tr (U + U^{-1})/2\rangle$ is plotted.
The data from complex Langevin dynamics with $\mu^2 \ge 0$ is consistent with those from real Langevin dynamics with $\mu^2\le 0$ in both
phases, including the critical region,
suggesting that complex Langevin dynamics is working correctly here. This is in contrast to the case of the XY model recently studied
using complex Langevin dynamics, where correct results were obtained in only part of the phase diagram~\cite{Aarts:2010aq}. 

At larger values of $\mu^2$, one can no longer rely on analytical continuation to justify the results. Instead, we assess them
using the criteria developed above and compute $\langle LO\rangle$ for $O = \Tr U$.
The outcome is shown in Figure \ref{fig:cc} for real chemical potential (left) and also in the phase-quenched theory (right), using
$\mu=3$ and $3.5$. In the phase-quenched theory, the action is real and we use real Langevin dynamics.
The figures indicate that at nonzero stepsize the criterion is not satisfied, both for real and complex Langevin dynamics,
but that in the limit $\epsilon \to 0$ it is satisfied.

We find therefore that in this model the test is passed successfully. Moreover, finite stepsize corrections can be quantified by the deviation
of $\langle LO\rangle$ from zero.
We are currently extending this analysis and have implemented a higher-order algorithm to eliminate the effects of finite stepsize~\cite{aarts-james}.

\section{Conclusions}
Complex Langevin dynamics can in principle be applied where the sign problem prevents the use of importance sampling. An analysis
of the long-time evolution of real and complex measures using the Fokker-Planck equation shows that the correct stationary solution
exists. However, if the decay of the distribution $P(x,y;t)$ is insufficient to allow integration by parts without boundary terms,
convergence to the wrong limit can occur. In practice, this can be diagnosed by testing the condition that $\langle LO\rangle = 0$
for a suitably large set of observables in the limit of vanishing stepsize.
In the case that the criterion is not satisfied, complex Langevin dynamics fails.

An analysis of the SU(3) spin model shows that the criterion for correctness is satisfied, justifying the claim that complex
Langevin dynamics works correctly with this model. This is corroborated by analytic continuation from imaginary chemical potential
in the region of small chemical potential $\left|\mu^2\right| \lesssim 1$, which shows that data for observables even in $\mu$ are continuous between
results from a real action when $\mu^2 \le 0$ and a complex action with $\mu^2 > 0$. 
A further and more detailed study of the criterion and stepsize dependence is in progress~\cite{aarts-james}.

\section*{Acknowledgments}
This work is supported by STFC.

\end{document}